\documentclass[twocolumn, pra, showpacs, aps, superscriptaddress]{revtex4-1}

\usepackage{amsmath}
\usepackage{amsfonts}
\usepackage{bbold}

\usepackage{graphicx}
\usepackage{ae,aecompl}
\usepackage{bm}
\usepackage{textcomp}
\usepackage{xcolor}

\usepackage{wasysym}
\usepackage{hyperref}

\begin{document}


\date{\today}
\title{Dating glacier ice of the last millennium by quantum technology}

\author{Z. Feng}
\email{iceArTTA@matterwave.de}
\affiliation{Kirchhoff-Institute for Physics, Heidelberg University}
\author{P. Bohleber}
\affiliation{Institute of Environmental Physics, Heidelberg University}
\affiliation{Institute for Interdisciplinary Mountain Research, {\"O}AW Innsbruck}
\author{S. Ebser}
\affiliation{Kirchhoff-Institute for Physics, Heidelberg University}
\author{L. Ringena}
\affiliation{Kirchhoff-Institute for Physics, Heidelberg University}
\author{M. Schmidt}
\affiliation{Kirchhoff-Institute for Physics, Heidelberg University}
\affiliation{Institute of Environmental Physics, Heidelberg University}
\author{A. Kersting}
\affiliation{Institute of Environmental Physics, Heidelberg University}
\author{P. Hopkins}
\affiliation{Institute of Environmental Physics, Heidelberg University}
\author{H. Hoffmann}
\affiliation{Institute of Environmental Physics, Heidelberg University}
\affiliation{Alfred Wegener Institute, Helmholtz Center for Polar and Marine Research, Bremerhaven}
\author{A. Fischer}
\affiliation{Institute for Interdisciplinary Mountain Research, {\"O}AW Innsbruck}
\author{W. Aeschbach}
\affiliation{Institute of Environmental Physics, Heidelberg University}
\affiliation{Heidelberg Center for the Environment, Heidelberg University}
\author{M.~K. Oberthaler}
\affiliation{Kirchhoff-Institute for Physics, Heidelberg University}

\begin{abstract}
Radiometric dating with $^{39}$Ar covers a unique timespan and offers key advances in interpreting environmental archives of the last millennium. 
Although this tracer has been acknowledged for decades, studies so far have been limited by the low abundance and radioactivity, thus requiring huge sample sizes.
Atom Trap Trace Analysis, an application of techniques from quantum physics such as laser cooling and trapping, allows to reduce the sample volume by several orders of magnitude, compared to conventional techniques. 
Here we show that the adaptation of this method to $^{39}$Ar is now available for glaciological applications, by demonstrating the first Argon Trap Trace Analysis (ArTTA) dating of alpine glacier ice samples.
Ice blocks as small as a few kilograms are sufficient and have been obtained at two artificial glacier caves. 
Importantly, both sites offer direct access to the stratigraphy at the glacier base and validation against existing age constraints. 
The ice blocks obtained at Chli Titlis glacier in $3030$~m~asl (Swiss Alps) have been dated by state-of-the-art micro-radiocarbon analysis in a previous study. 
The unique finding of a bark fragment and a larch needle within the ice of Schaufelferner glacier in $2870$~m~asl (Stubai Alps, Austria) allows for conventional radiocarbon dating. 
At both sites, results of $^{39}$Ar dating match the existing age information based on radiocarbon dating and visual stratigraphy.
With our results, we establish Argon Trap Trace Analysis as the key to decipher so far untapped glacier archives of the last millennium.
\end{abstract}
\maketitle 

\section{Introduction} 
Non-polar glaciers are dynamic archives of environmental change, covering altitudes where other climate records are sparse. 
We find a unique situation in the European Alps comprising the densest network of long instrumental climate observations and anthropogenic emission sources in immediate vicinity of glaciers and other climate archives. 
Only few glaciers of the highest summit regions, typically above $4000$~m~asl, archive snow and thus past climate signals on a quasi-continuous basis. 
Glaciers at summit locations of lower altitudes are more abundant, but have only recently been investigated for their potential as climate archives \cite{haeberli2004characteristics}. 
If their ice is frozen to bedrock, the age structure of these glaciers contains valuable information about the climatic conditions at the onset of formation.
Because periods without net accumulation or even prolonged mass loss can occur at these glaciers, their stratigraphy does not include layers of every single year, making dating by annual layer counting impossible. 
Hence, age constraints can only be obtained from radiometric methods yielding an absolute age information.

According to the age range accessible by their half-lives, $^{3}$H and $^{210}$Pb are established tools to constrain the age of glacier ice within the last $100$~years \cite{gaggeler1983210}. 
Micro-radiocarbon techniques building on analysis of particulate organic carbon extracted from glacier ice are now available for dating ice samples older than roughly $1000$~years \cite{hoffmann2018new,uglietti2016radiocarbon}, but for ages younger than this, the $^{14}$C technique is hampered by ambiguities in the calibration curve as well as limited sample size \cite{ramsey1995radiocarbon,reimer2013intcal13}.
Likewise as for glaciers in many non-polar mountain ranges, a substantial part of the ice volume in the European Alps may not reach maximum ages that fall within the range of the $^{14}$C technique. 
Even if the lowermost layer of a glacier can be dated by radiocarbon methods, the larger portion of the stratigraphy is likely to be substantially younger and hence not suitable for the application of $^{14}$C.
As a result, there is an immediate demand in the glaciological community for radiometric dating of glacier ice within the age range of $100-1000$~years.
A particular example is the Little Ice Age period from late 13th to mid 19th century.
For climate science, this is a key period for understanding climate variability, and well suited for model calibration as instrumental and historical climate data is available for cross validation of climate proxies and model results, in the Alps from about $1500$~CE onwards \cite{dobrovolny2010monthly}. 

A powerful tracer lies within the air bubbles enclosed in glacier ice: the rare isotope $^{39}$Ar.
It is a unique tracer with a half-life of $269$ years \cite{Stoenner1325}, hence matching the time span of $100-1000$~years. 
The key to decipher the information stored in climate archives of glaciers are small ice samples of the order of kilograms to achieve the required spatial and thus temporal resolution. 
However, the entrapped gas content and low $^{39}$Ar abundance leads to only $2,000 - 20,000$ $^{39}$Ar atoms contained in one kilogram of modern ice, putting strong challenges in quantitative detection. 

A much larger number of atoms is required for classical analysis by low level decay counting implying the need for significantly larger sample sizes in the order of tons \cite{oeschger1976ice, LOOSLI198351}. 
For this reason, environmental routine measurements of $^{39}$Ar have until today mainly been confined to groundwater reservoirs where nearly unlimited sampling is possible, e.g. in \cite{alvarado2007groundwater}. 
A strong reduction in the required sample size has become feasible recently by the method of Atom Trap Trace Analysis. 
This technique utilizes the isotopic shift in optical resonance frequency to capture single atoms of the desired isotopic species. 
The required multi-photon scattering for this process yields perfect selectivity. 
It was originally developed for the isotope $^{81}$Kr \cite{Chen1139} and has been applied in several studies \cite{LU2014196}. Dating old antarctic ice has been demonstrated using samples of several hundred kilograms \cite{buizert2014blueice}. 

The adaptation of this method to $^{39}$Ar, which we refer to as Argon Trap Trace Analysis (ArTTA), and its application to environmental studies has been demonstrated with large groundwater \cite{ritterbusch2014groundwater} and recently with small ocean samples \cite{ebser2018ocean}.  
ArTTA is thus the door-opener for broad application of radio-argon dating in research fields as glaciology, that have so far been excluded, due to sample size requirements \cite{lu2014argondating}.

\section{Site description and sample selection} 
\begin{figure*}[t]
\centering
\includegraphics[width=\linewidth]{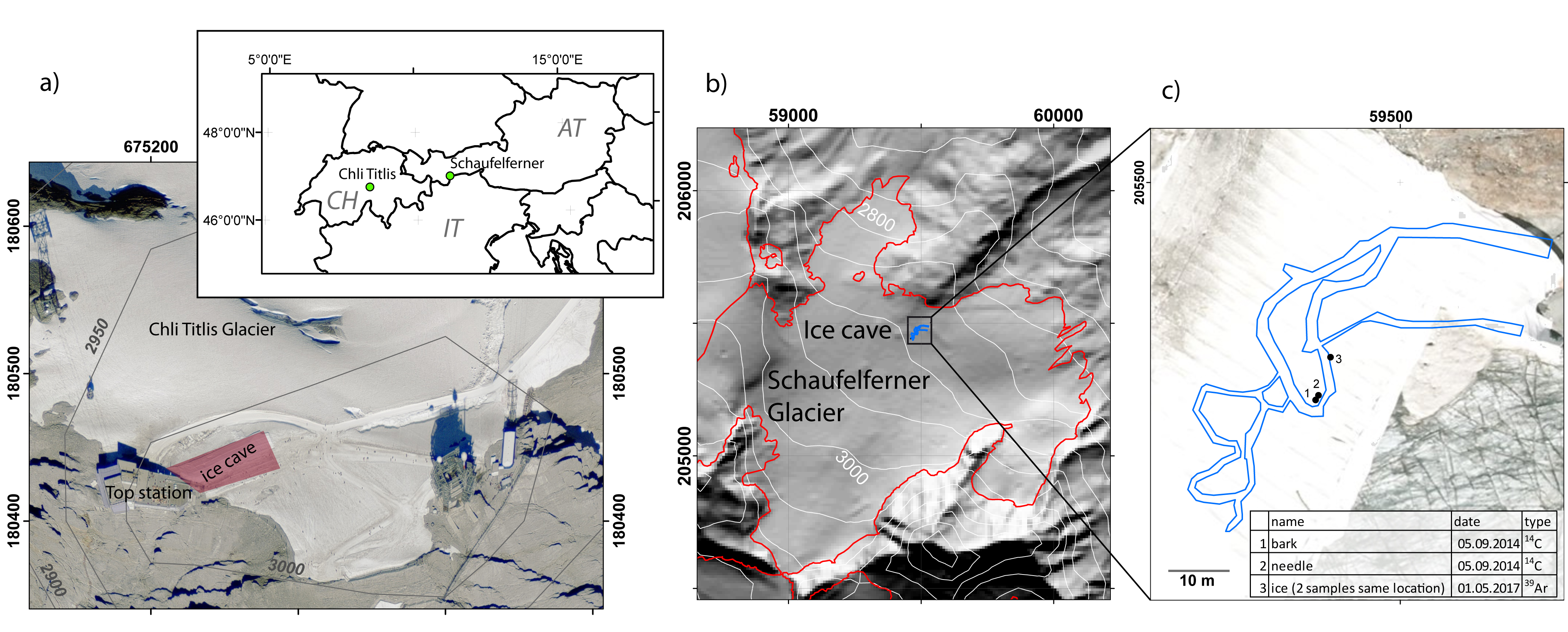}
\caption{Site maps of a) Chli Titlis glacier cave and b) Schaufelferner glacier cave with c) schematic view of the Schaufelferner ice tunnel and sample locations. Due to the summit location, the ice at the Chli Titlis cave is nearly stagnant, showing horizontal layers allowing straightforward sampling and relative age control via stratigraphy, i.e. older ages at greater depth. The Schaufelferner glacier cave is located downstream of the summit and has undergone substantial ice flow. Inclined layers undisturbed by folding are visible within the cave. The GPS coordinates for Chli Titlis and Schaufelferner are reported in the Swiss grid and Gauss-Krueger system, respectively.}
\label{fig:sampling_sites}
\end{figure*}

For the purpose of this study we selected two glacier sites (see Figure \ref{fig:sampling_sites}) offering artificial glacier caves, which make highly controlled sampling of suitable sample sizes for ArTTA possible. 
Additionally, the glacier caves provide direct access to the internal glacier stratigraphy and thus relative age control.
The cornice-type summit at Chli Titlis ($3030$~m~asl, Central Switzerland) holds the glacier on its north-facing slope, with a tunnel dug for touristic purposes around $100$~m into the ice along bedrock starting at the cable car station \cite{lorrain1990climatic}.
Schaufelferner in the Stubai Alps (Austria) is part of the Stubai glacier ski resort and covers altitudes from $3270$~m~asl to $2810$~m~asl. 
The glacier tongue of Schaufelferner is divided into two parts at the Eisgrat nunatak. 
At the orographic right side of the nunatak, the glacier cave was drilled in 2013 CE, close to the cable car station. 

The visual stratigraphy of the ice at both sites shows abundant bubble-rich layers of white appearance, with occasional transparent, nearly bubble-free layers originating from refrozen meltwater. 
At Chli Titlis and Schaufelferner, ice flow velocities inside the caves (i.e. near bedrock) are close to zero meaning that the sampling locations have remained nearly unchanged from the beginning of the samplings in 2014 CE up to 2018 CE. 
At Chli Titlis cave, ice temperatures are below zero throughout the year aided by artificial cooling. 
At Schaufelferner cave initial englacial temperature measurements also revealed ice frozen to bedrock. 
Continuous temperature monitoring is currently underway to further investigate the spatial distribution of seasonal variability of englacial temperatures. 
The glacier surface above both Chli Titlis and Schaufelferner glacier caves is subject to rapid shrinking and seasonally covered by sheets to minimize summer ice melt. 
The clear evidence of prolonged negative mass balance means that the unique paleo-climate information of these sites may not be preserved much longer. 
Both sites have been the subject of previous glaciological investigations, but differ primarily by being located within the nearly stagnant summit region (Chli Titlis) vs. a site having undergone substantial ice flow (Schaufelferner). 
Accordingly, our sampling strategy was guided by the idea to obtain two samples (i) of neighboring layers significantly different in age (Chli Titlis) and (ii) within a single layer of the same age (Schaufelferner). 

\begin{figure}
\centering
\includegraphics[width=\linewidth]{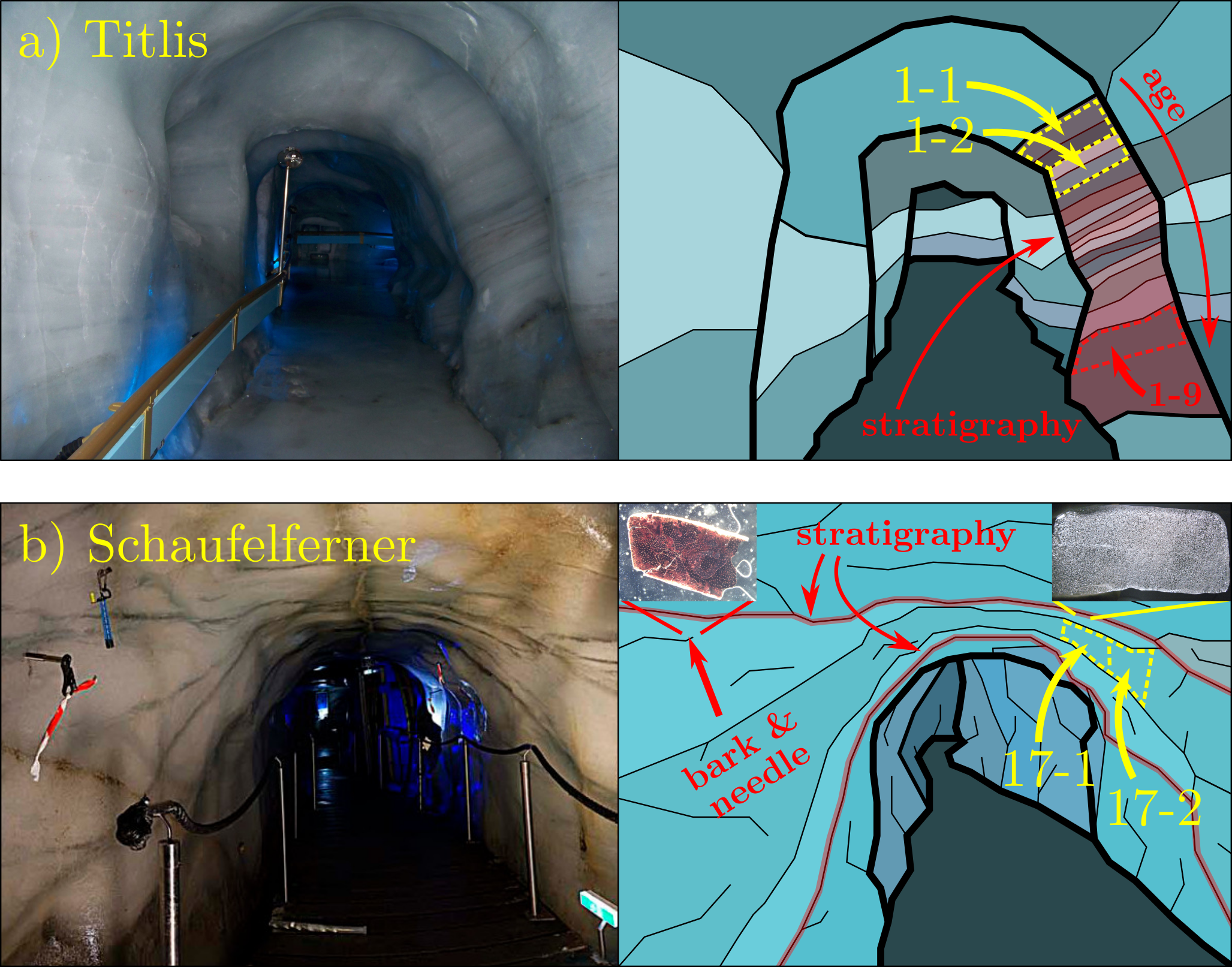}
\caption{a) The sampling site at the summit of Chli Titlis. The wall in the cave shows a distinct horizontal layering, hence no evidence of layer folding. The $^{14}$C age constraints reveal a strong vertical age gradient within the sampled depths \cite{bohleber2018investigating}. The two uppermost blocks (1-1 and 1-2) have been analyzed for $^{39}$Ar. The results of $527^{+119}_{-156}$ years before 2018 CE for the uppermost block 1-1 and $1126^{+1286}_{-273}$ years before 2018 CE for block 1-2 are realistic in view of existing age evidence provided by visual stratigraphy and $^{14}$C dating results. b) Schaufelferner glacier cave. A bark particle (inset) and a larch needle have been extracted from the wall and allow for macroscopic $^{14}$C dating (see \cite{hoffmann2016micro} for more details). The ice blocks of suitable size used for $^{39}$Ar dating have been taken in a later campaign, a few meters apart from the original location. Great care has been applied to select a layer for $^{39}$Ar as close as possible to the layer comprising the organic objects. The $^{39}$Ar dating results of $193^{+53}_{-55}$ and $198^{+60}_{-64}$ years before 2018 CE agree with $^{14}$C findings (see main text).}
\label{fig:glacier_caves}
\end{figure}

At Chli Titlis, ice blocks of approximately $4$~kg were cut out by chain saw. 
The outermost layer exposed to the tunnel was removed prior to collecting sample blocks to avoid contamination due to cracks or melt water. 
Micro-radiocarbon dating revealed a strong vertical age gradient (see Figure~\ref{fig:glacier_caves}a). 
For instance, $^{14}$C ages of a profile sampling $1.9$~m of the lowermost ice of the glacier range from $1246$ (block~1-2) to $3138$ (block~$1-9$) years before 2018 CE. 
A maximum age of $5387$~years before 2018 CE was found at the inner part of the glacier cave.
For later direct comparison to $^{39}$Ar, all calibrated radiocarbon ages have been adjusted to refer to the year 2018 CE as present.
Further details regarding the sampling methods and the site characteristics of the Chli Titlis glacier cave can be found in \cite{bohleber2018investigating}. 
Based on the $^{14}$C age constraints, we selected the two youngest ice blocks for our comparison with $^{39}$Ar, blocks 1-1 and 1-2. 

At Schaufelferner, the ice in the cave has flowed downwards from the top yielding tilted layering without folding.
A unique feature of this site is the rare finding of two macroscopic particles of organic origin, a bark and a larch needle, inside the ice (Figure~\ref{fig:glacier_caves}b). 
Both objects have been radiocarbon dated in a previous campaign. 
The age range obtained from the bark and needle is so far considered the best representation of the actual age of this ice layer \cite{hoffmann2016micro}. 
For $^{39}$Ar analysis, two adjacent blocks were obtained by chain saw within one layer at the location indicated in Figures~\ref{fig:sampling_sites}c and \ref{fig:glacier_caves}b. 
The according location is at roughly $4$~m horizontal distance from the macroscopic organic particles.
The $^{39}$Ar sampling location has been chosen as close as possible to the layer containing the organic objects while permitting to obtain kg-size ice blocks of undisturbed, bubble-rich ice. 
The layers sampled for $^{39}$Ar and the organic particles are stratigraphically in close proximity, in spite of the horizontal distance (as indicated in Figure~\ref{fig:sampling_sites}c.). 

\section{Discussion of ArTTA results}
\begin{table*}[t]
\centering
\caption{Results of Argon Trap Trace Analysis of glacier ice samples}
\begin{tabular}{llllllll}
Sample Name & Concentration [pmAr] & Argon Age [a]  & Counted Atoms & Est. Background & Time [h] & Size [mL STP] & Weight [kg]\\
\hline
Titlis 1-1 & $25.7^{+9.2}_{-8.5}$ & $527^{+119}_{-156}$ & 50 & 26 & 20.00 & 0.5 & 4.3\\

Titlis 1-2 & $5.5^{+5.6}_{-5.3}$ & $1126^{+1286}_{-273}$ & 31 & 25 & 20.00 & 0.6 & 4.2\\

Schaufelferner A & $60.8^{+8.9}_{-8.0}$ & $193^{+53}_{-55}$ & 92 & 13 & 22.30 & 1.7 & 6.7 \\

Schaufelferner B & $60.0^{+10.1}_{-9.1}$ & $198^{+60}_{-64}$ & 66 & 10 & 19.50 & 1.4 & 5.9\\
\hline
\label{tab:Results}
\end{tabular}
\end{table*}

\begin{figure*}
\centering
\includegraphics[width=\linewidth]{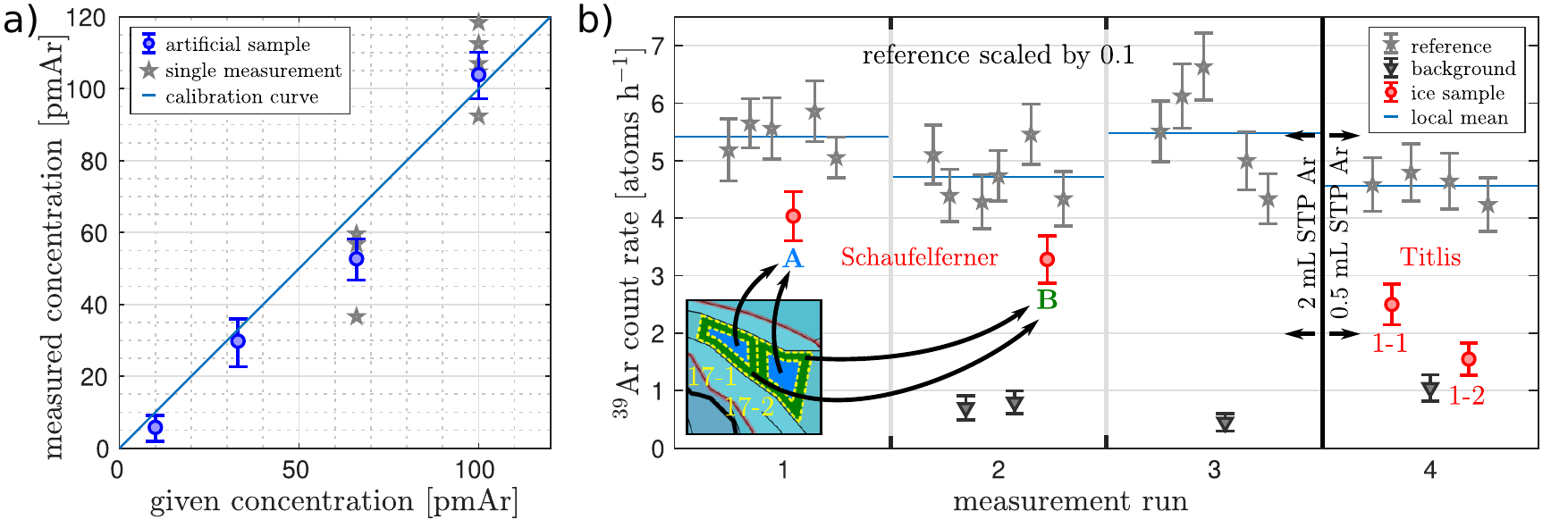}
\caption{a) Calibration of ArTTA with known samples. Three artificial concentrations have been produced by mixing an $^{39}$Ar-free sample with modern argon. The $10$~pmAr and $33$~pmAr samples have been measured once whereas the mean values of three $66$~pmAr and five $100$~pmAr measurements (stars) are given (see \cite{ebser2018phd} for more details). The analysis is consistent within the given $1\sigma$ confidence interval and confirms the possible dating range of ArTTA. An ideal calibration line is shown as guide to the eye. b) Sequence of ice analysis. In weekly measurement runs, the local mean values of several references (stars) are used to infer the concentration of the samples. Background can be neglected for these measurements due to their $10 \times$ modern concentration and short measurement time of $2$~h. Contrarily, the counts of sample measurements (circles) have to be corrected for background. This parameter is obtained by background measurements with $^{39}$Ar-free samples (triangles). To verify the necessary effort for ice sample decontamination, the similar ice blocks 17-1 and 17-2 from Schaufelferner have been prepared by two different approaches to yield samples A and B (inset).}
\label{fig:calibration_timeline}
\end{figure*}

In order to ensure reliable performance of the ArTTA setup for the necessarily small sample sizes, artificial samples of known concentration have been prepared and analyzed.
The results are shown in Figure \ref{fig:calibration_timeline}a (see Table~S1 and \cite{ebser2018phd} for more details).
Samples with $2$~mL STP argon and concentrations of $66$~pmAr, $33$~pmAr and $10$~pmAr have been produced by mixing modern argon with an $^{39}$Ar-free sample provided by the dark matter search collaboration, Darkside \cite{darkside2016deadargon}.
These results confirm the capability for reliable dating by ArTTA within the age range of $100-1000$~years before 2018 CE. 

The glacier ice samples were analyzed by ArTTA, Table \ref{tab:Results} provides an overview of all important parameters and measurement results (see Table~S2 for more details).
Notably, due to very low air content, the samples of the two ice blocks of Chli Titlis yielded as little as $0.5$~mL STP argon. 
However, count rates were still significantly above background (see method section) and allowed for reliable measurements. 

The two adjacent ice blocks 17-1 and 17-2 from Schaufelferner were prepared for ArTTA following two different approaches, yielding samples A and B. (see inset Figure~\ref{fig:calibration_timeline}b).
For the Schaufelferner A sample, the core parts of both blocks were obtained by carefully removing a cm-thick layer of every surface exposed to the atmosphere, followed by additional scraping of the cut surface using a microtome.
The Schaufelferner B sample combines these removed surface layers and was only cleaned of sections evidently containing refrozen (i.e. clear, bubble-free) ice.
The fact that the results of sample A and B are not significantly different indicates that contamination by modern air is unlikely and that simple chain saw sampling is indeed possible as it does not seem to spoil the radioargon dating. 

Regarding the actual dating results of the ArTTA analyses, we performed a comparison with age constraints by $^{14}$C, carefully taking into account (i) the different $^{14}$C sample types (macroscopic vs. microscopic) and (ii) relative age information provided by the visible stratigraphy in the two caves.

The two ice blocks of Chli Titlis are dated at the far old end of the time span accessible to $^{39}$Ar dating. 
Block 1-1 is $^{39}$Ar dated at $527^{+119}_{-156}$ years before 2018 CE where the error is mainly given by finite counting statistics. 
The adjacent block below, block 1-2, is at least $170$ years older, with a best-estimate argon age of $1126^{+1286}_{-273}$ years before 2018 CE. 
Constraints on the age by $^{14}$C dating exist only for block 1-2, where we obtain $1246-1378$ years before 2018 CE (see Figure~S1). 
It is important to note that no macroscopic organic fragments were found at Chli Titlis, and all $^{14}$C analyses had to be performed on microscopic particulate organic carbon, for which known biases towards older ages can exist. 
Following the discussion of potential reservoir effects in \cite{hoffmann2018new}, the $^{14}$C age is thus regarded as upper age limit \cite{bohleber2018investigating}.
Additionally, important relative age control is provided by the evident near-horizontal layering, i.e. older ice is located at greater depth (see Figure \ref{fig:glacier_caves}a). 
The $^{14}$C results revealed that age differences of several hundred years occur at close range, even between two adjacent blocks. 
The large age gradient in the lowermost ice layers is not necessarily connected to layer thinning by deformation but can instead result from past hiatuses in glacier growth or intermittent melting periods \cite{bohleber2018investigating}.
In this context, the $^{39}$Ar ages perfectly match what is known to-date about the Chli Titlis glacier cave, namely (i) that age differences of up to several hundred years can occur even between two adjacent blocks of $\sim 20$~cm height, (ii) the $^{39}$Ar age of $1126^{+1286}_{-273}$ years of block 1-2 is a match against the lower carbon age limit of 1246 years before 2018 CE. The $^{39}$Ar age for block 1-1 of $527^{+119}_{-156}$ years adds new information and is consistent with the already known vertical age gradient connected to the intermittent ice build-up process.

At Schaufelferner glacier cave, the $^{39}$Ar results of samples A and B are consistent and reveal a layer age of $193^{+53}_{-55}$ and $198^{+60}_{-64}$ years before 2018 CE, respectively. 
For comparison with existing age constraints, it is important to note that the bark particle and larch needle were dated by conventional macroscopic $^{14}$C analysis. 
This means that reservoir effects related to the various microscopic $^{14}$C fractions are negligible (cf. discussion of Chli Titlis above). 

The uncertainty in radiocarbon dating is caused by ambiguities in calibration of $^{14}$C ages within the respective time period. 
Using OxCal v4.3.2 \cite{bronkoxcal} and the IntCal13 atmospheric calibration curve \cite{reimer2013intcal13}, the most likely age ranges assigned to the bark particle and the larch needle are $375-532$ and $505-632$ years before 2018 CE at $68\%$ and $48\%$ probability (see Figures~S2 and S3), respectively \cite{hoffmann2016micro}. 
However, it should be mentioned that, consistently for both objects, the calibration also reveals a small but non-zero probability for ages around $230$ years before 2018 CE, which would be a direct match with the $^{39}$Ar dating results. 
Regardless of the calibration issue, the $^{14}$C age range of the macroscopic organic particles has to be considered as an upper estimate of the age of the glacier ice. 
This is due to the potential delay in deposition on the glacier surface after creation of the macroscopic organic fragments, e.g. death of the respective tree. 
No such age-offset exists for $^{39}$Ar. 
In contrast to what is known from polar studies regarding systematic age-offsets between the enclosed air and ambient ice matrix \cite{sowers1992delta15n}, no significant effect in this regard is expected within the $^{39}$Ar dating uncertainty. 
This is due to the rapid formation of ice at our study sites, typically of the order of a decade or less \cite{ambach1966analysis}.
Taking the most probable calibrated $^{14}$C range for comparison with $^{39}$Ar results in a difference of at least $85$ years between the $^{14}$C and the $^{39}$Ar ages. 
This age difference is well within what can be explained based on glaciological considerations. 
The $^{39}$Ar and $^{14}$C dating results refer to neighboring layers.
Likewise as for Chli Titlis, an age difference of the order of decades is reasonable at Schaufelferner due to hiatuses and melting periods. 

Based on our results we find no evidence of contamination related to our sampling method by chain saw. 
Thus sampling is comparatively simple allowing to obtain blocks of convenient size of $\sim 5$~kg. 
The results of Chli Titlis and Schaufelferner are the first demonstration of conclusive $^{39}$Ar dating of glacier ice with samples containing less than $2$ mL STP argon. 
In concert with evidence provided by the visual stratigraphy, the comparison with $^{14}$C age constraints corroborates the ArTTA age dating method, both for its mid- (Schaufelferner) and far-end (Chli Titlis) age range. 
At Chli Titlis, the $^{39}$Ar dating results show an age range and a vertical age gradient that reproduce and extend earlier findings obtained from $^{14}$C. 
For Schaufelferner, the most likely age ranges assigned by $^{14}$C dating of macroscopic organic objects are determined systematically older than the $^{39}$Ar dates, but stay within a range that can be explained based on glaciological considerations. 
Furthermore, both $^{39}$Ar and $^{14}$C ages indicate that the ice at Schaufelferner is likely a remnant of the 1850 glacier maximum. 
The $^{39}$Ar dating tool provides a more reliable age constraint in this case, considering the ambiguity associated with the $^{14}$C age calibration. 
Thus, the ice in the Schaufelferner cave originates from the Little Ice Age glacier advance.

\section{The future of glacier ice dating with ArTTA}
Since glaciers at other non-polar mountain ranges, e.g. Central Asia, Himalaya or Andes, are not much different from the Alps regarding their glaciological characteristics, e.g. size, accumulation and englacial temperature, the impact of this study goes beyond the Alpine realm. 
However, the European Alps can provide a unique combination of (i) glaciers featuring expected age ranges suitable for $^{39}$Ar, (ii) access to kg-size ice samples at low cost through excellent infrastructure and (iii) availability of multi-proxy reconstructions of Holocene climate. With the introduction of the ArTTA ice dating technique, we can retrieve so far untapped paleoclimate records and validate them in the nexus of European archives.
The scientific main potential of $^{39}$Ar dating of glacier ice is the interpretation of the ice layers formed during the last millennium, in order to reveal the past history of summit glacier growth in the Alps. 
This history comprises a fundamental gap in knowledge in the context of the highly complex climate patterns during the core period of the Little Ice Age. 
Even the analysis of ice dynamics during the last centuries can be done with $^{39}$Ar by sampling along the central flow line of a glacier. 
Past ice dynamics can be considered an important, but fairly unknown parameter governing the reaction of glaciers to climate change.
Accordingly, developing the full potential of dating by $^{39}$Ar will be key to opening up new avenues in paleo-climatic research.
As current warming conditions pose an immediate threat to losing this precious information, the novel dating tool of $^{39}$Ar is arriving just-in-time in modern glaciological research. 
It also generates a broader impact in the field of Holocene climate science. 
In this sense, the $^{39}$Ar dating technology for glacier ice has the potential to yield a major scientific advance comparable to the introduction of radiocarbon or surface exposure dating. 

\section{Material and Methods}

\subsection*{Ice processing and argon extraction}
The ice blocks were transported from the glaciers to a $-20$~$^\circ$C cold storage at Heidelberg University.
Prior to the extraction, they were cleaned by cutting off melted layers. 
Ice blocks of up to $8$~kg were then put into a $12.6$~L stainless steel container which was evacuated with a turbo molecular pump. 
The ice was melted and the gas was extracted by freezing it through a water trap onto a liquid nitrogen cooled activated charcoal trap. 

ArTTA is highly selective and thus immune to impurities due to other elements. 
Still, a high argon purity and yield was desired to maximize the counting efficiency and minimize the sample amount. 
For this purpose, a specific argon purification system had been designed. 
The gas composition was analyzed with a quadrupole mass spectrometer before the gas was transferred to a $900$~$^\circ$C titanium sponge getter. 
With that, all gases were removed except for noble gases and hydrogen. 
At a second titanium sponge getter at room temperature the hydrogen was adsorbed and the remaining gas fraction, consisting of $> 99$~\% argon, was captured on a charcoal trap and transported to the ArTTA setup. 
With this system, an ice sample was processed within 4 hours with an argon recovery rate of $ >98 \text{ \%}$.

\subsection*{ArTTA setup}

\begin{figure}[h]
\centering
\includegraphics[width=\linewidth]{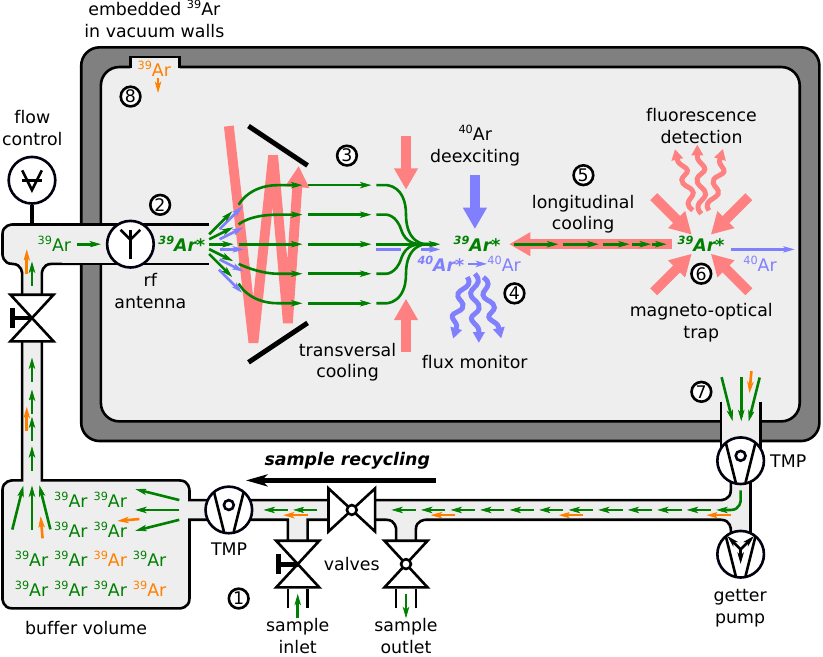}
\caption{Setup of Argon Trap Trace Analysis. (1) The sample is compressed into a buffer to compensate for different sample volumes. (2) Argon is excited into a metastable state for convenient optical transitions at $812$~nm. (3) Transversal laser cooling is applied to increase the $^{39}$Ar flux into the detection region. (4) $^{40}$Ar is deexcited to the ground state and provides a signal for flux monitoring. (5) Longitudinal laser cooling slows the thermal atoms down to tens of meters per second. (6) Single $^{39}$Ar atoms are captured in a magneto-optical trap and the fluorescence is monitored by a photo diode. (7) Several turbo-molecular pumps (TMP) collect the gas from the apparatus. The sample is cleaned by a getter pump and is re-compressed to the buffer volume. (8) Enriched $^{39}$Ar out-gasses from the vacuum walls and contaminates the sample. This effect scales with volume and limits the analysis for smaller samples.}
\label{fig:artta_setup}
\end{figure}

A simplified scheme of the ArTTA apparatus is shown in Figure \ref{fig:artta_setup}. 
The purified Ar gas of the ice sample is first compressed into a buffer volume to compensate for different sample sizes while obtaining a constant flow into the main apparatus. 
Laser cooling is building on strong closed dipole transitions which are available for metastable argon (Ar*). Thus, metastable argon is produced in an RF-discharge source. 
Liquid-nitrogen cooling is applied to reduce the initial velocity of the atoms.

The flux into the small detection region is increased by two transversal laser cooling stages. 
The first stage collects the divergent atoms from the effusive source and collimates them to a beam, whereas the second stage compresses the beam. Subsequently, the longitudinal velocity is reduced from thermal velocities to a few meters per second by a Zeeman slower.
Single $^{39}$Ar atoms are finally captured and detected in a magneto-optical trap, thus guaranteeing perfect selectivity by millions of resonant photon scattering processes in a spatially confined region.
To reduce the off-resonant scattering of the huge isotopic background of abundant $^{40}$Ar in the detection region, this isotope is selectively de-excited from the metastable state to the ground-state by an additional quench laser. 
The fluorescence of this process provides a direct signal for flux monitoring. 

Several turbo-molecular pumps realize the necessary ultra-high vacuum in the apparatus. 
Their collected gas is cleaned with a non-evaporative getter by removing any non-noble gas contributions.
The restored gas is compressed into the buffer volume again thus enabling full recycling of the sample. 
With this procedure the required sample size is as low as $0.5$~mL~STP. 

The sample size of the current ArTTA setup is limited by out-gassing of embedded argon enriched in $^{39}$Ar.
The contribution due to this contamination is dependent on sample size, actual concentration and measurement time. 
Around $10$~pmAr is expected for a volume of $2$~mL STP in a $20$~h measurement.

\subsection*{Measurement procedure}

ArTTA is currently capable of analyzing one sample per day with relative accuracy dependent on the actual concentration. 
A full measurement cycle starts with $\sim 20\ \text{h}$ of sample measurement. 
This is directly followed by $\sim 2\ \text{h}$ of referencing to an artificial sample with $10\times$ enriched $^{39}$Ar compared to modern concentration. 
The apparatus is flushed with a krypton discharge while refilling the liquid nitrogen reservoir to remove any frozen content on the source and reduce cross sample contamination. 
Each sample is framed by at least two reference measurements and it is this temporal local referencing which makes the measurement robust against long-term drifts (see \cite{ebser2018phd} for more details).
\subsection*{Data processing}

In order to infer the sample concentration from the number of atoms detected, careful estimate of the background is necessary.
The contribution due to embedded argon enriched in $^{39}$Ar can be determined by background measurements with $^{39}$Ar-free samples \cite{darkside2016deadargon}. 
This long-term memory effect is described by a constant out-gassing of $^{39}$Ar yielding a time-dependent concentration

\begin{align*}
c(t) &= c_{sample} + \frac{a_{out}}{V_{sample}} t
\end{align*}

with sample concentration $c_{sample}$, sample volume $V_{sample}$ and $^{39}$Ar out-gassing rate $a_{out}$. 
By integration over time, the detected total atom number $N_{tot}$ is given by

\begin{align*}
N_{tot} &= \left[c_{sample} t + \frac{1}{2} \frac{a_{out}}{V_{sample}} t^2 \right] \cdot \lambda_0
\end{align*}

with $\lambda_0$ describing the count rate of a sample with modern concentration. This parameter is inferred from measurements of reference samples with $10\times$ modern $^{39}$Ar concentration. 

The model is used in a Bayesian analysis to obtain the probability density function for the sample concentration.
The reported concentrations are the extracted most probable values and the uncertainties correspond to the $1\sigma$ confidence interval containing $68.3$~\% of measurements (see \cite{ebser2018phd} for more details). 

Notably, the contribution to the detected atoms due to the background is increasing quadratically in time which limits the integration time and thus the accuracy of the inferred concentration.

\subsection*{Accuracy and limit}

\begin{figure}[h]
\centering
\includegraphics[width=\linewidth]{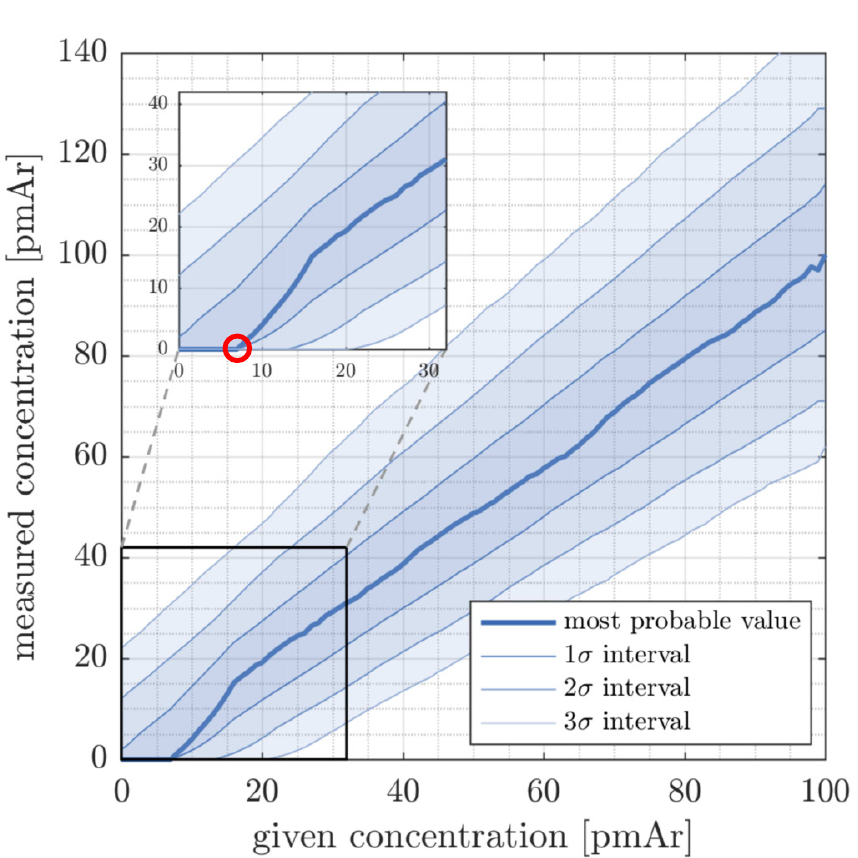}
\caption{Simulated accuracy for $0.5$ mL STP samples. For each given concentration, $10,000$ Monte-Carlo simulations of the here employed measurement routine have been performed using the experimentally determined parameters, i.e. out-gassing rate and modern count rate. This yields distributions of the most probable values of the inferred concentrations indicated by the mode and standard deviation. The range for measurable concentrations is exceeded as soon as the background contribution cannot be statistically distinguished anymore. For samples of $0.5$ mL STP, this is the case for most measurements below $8$~pmAr (inset).}
\label{fig:simulated_performance}
\end{figure}

In order to obtain the accuracy of our apparatus, we perform numerical simulations of the current measurement routine by using the experimentally determined parameters $\lambda_0$ and $^{39}$Ar out-gassing rate $a_{out}$. 
For each given concentration, the references, background as well as a total number of counted atoms have been generated numerically in $10,000$ Monte-Carlo simulations and analyzed in the same way as measured data.
The most probable values of the inferred concentrations are shown in Figure~\ref{fig:simulated_performance} by indicating the mode of this distribution as well as $1\sigma$, $2\sigma$ and $3\sigma$ intervals containing $68.3$~\%, $95.5$~\% and $99.7$~\% of the values respectively.
The contribution by the embedded contamination is especially dominant for small sample sizes and low concentrations. For the case of $0.5$~mL STP samples and below $8$~pmAr, the background cannot be statistically distinguished in most of the measurements, but a single measurements can yield concentrations significantly, i.e. $1\sigma$, above zero (see \cite{ebser2018phd} for more details).

\section{Acknowledgement}
This work was supported by the Deutsche Forschungsgemeinschaft (DFG, German Research Foundation), the FWF project Cold Ice (P29256-N36, sampling) and the European Research Council (ERC) under the European Union's Horizon 2020 research and innovation programme (Grant agreement No 694561). We further thank the operators of Titlis and Stubai ski resorts for their help in logistics.
  

\bibliographystyle{APS}


\widetext
\begin{center}
\textbf{\large Supplemental Materials: Dating glacier ice of the last millennium by quantum technology}
\end{center}

\setcounter{equation}{0}
\setcounter{figure}{0}
\setcounter{table}{0}
\setcounter{page}{1}
\makeatletter
\renewcommand{\theequation}{S\arabic{equation}}
\renewcommand{\thefigure}{S\arabic{figure}}
\renewcommand{\thetable}{S\arabic{table}}

\begin{figure}[h]
\centering
\includegraphics[width=\textwidth]{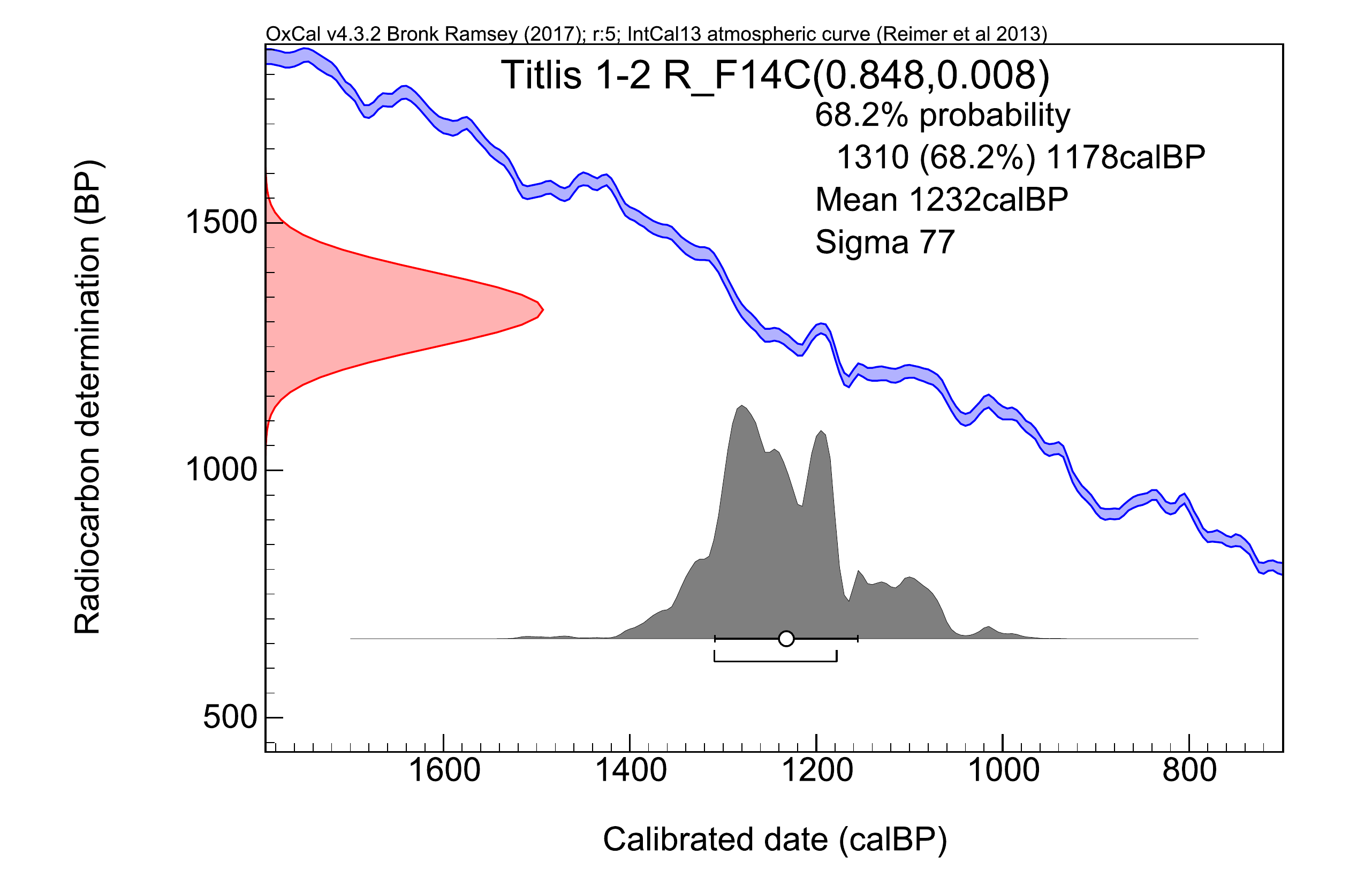}
\caption{$^{14}$C calibration curve of block 1-2 from Titlis glacier cave. The analysis has been performed on microscopic particulate organic carbon. The most probable age range is $1246-1378$ years before 2018 CE \citep{hoffmann2016micro}.}
\end{figure}

\clearpage
\newpage
\mbox{~}

\begin{figure}[h]
\centering
\includegraphics[width=\textwidth]{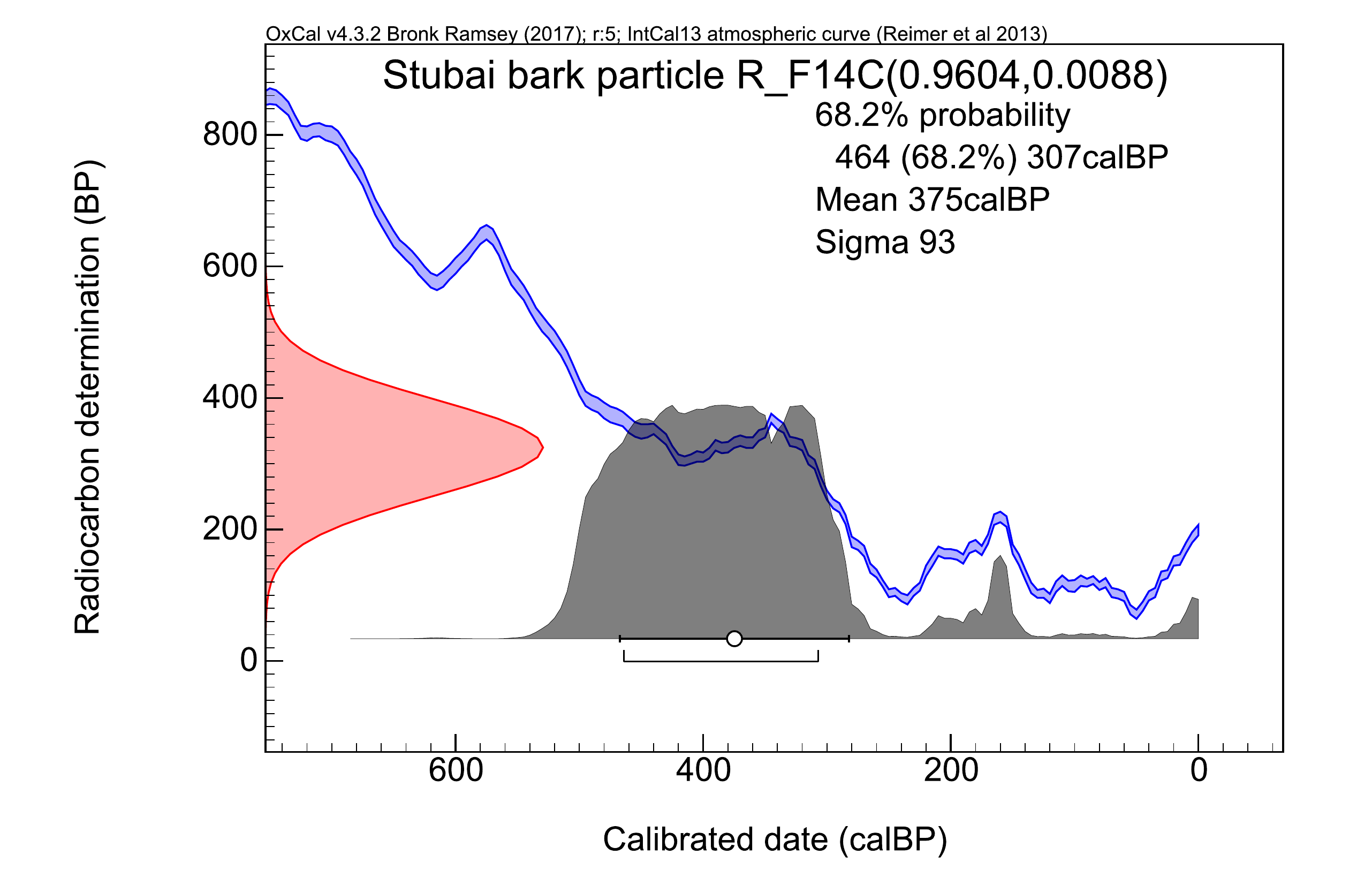}
\caption{$^{14}$C calibration curve of the bark findings in Schaufelferner glacier cave. The most probable age range is $375-532$ years before 2018 CE with $68\%$ probability. \citep{hoffmann2016micro}}
\end{figure}

\clearpage
\newpage
\mbox{~}

\begin{figure}[h]
\centering
\includegraphics[width=\textwidth]{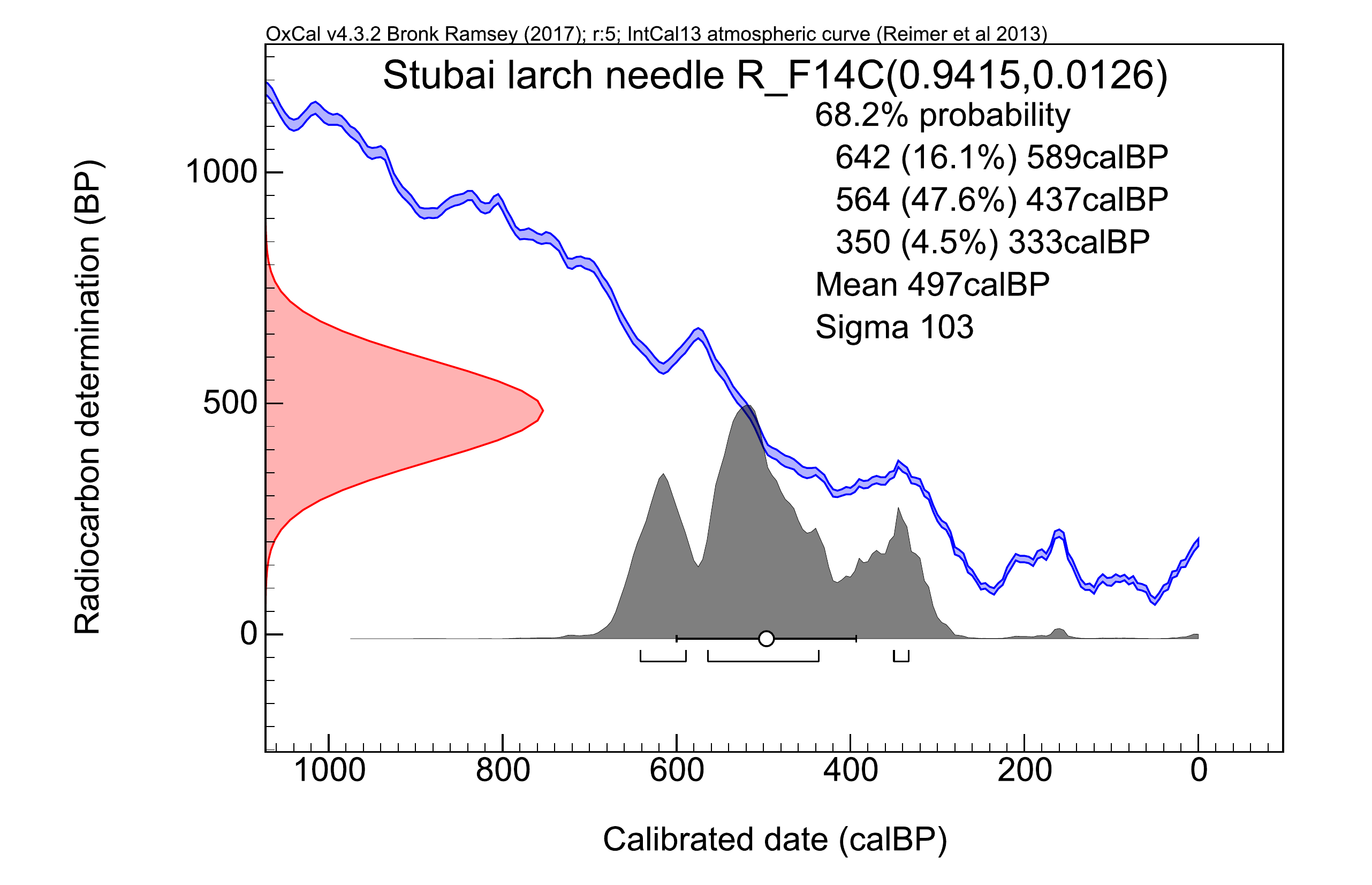}
\caption{$^{14}$C calibration curve of the needle findings in Schaufelferner glacier cave. The most probable age range is $505-632$ years before 2018 CE with $48\%$ probability. \citep{hoffmann2016micro}}
\end{figure}

\clearpage
\newpage
\mbox{~}

\begin{table}[h]
\centering
\caption{ArTTA calibration measurements. The calibration has been performed with samples of $2$~mL~STP~argon and the reported concentrations are the most probable value and $1\sigma$ confidence intervals \citep{ebser2018phd}. The dead argon is provided by the Darkside collaboration \citep{darkside2016deadargon}. This sample serves both as estimation of the background and sample of zero concentration.} 
\begin{tabular}{lrrrrrrr}
sample & number of measurements & concentration [pmAr] & total atoms detected & estimated background & total time [h]\\
\hline
dead & 5 & $ \leq 1.7 $ & 64 & 65 & 107.2\\
10\% mixture & 1 & $ 5.8^{+3.9}_{-3.3} $ & 18 & 11 & 20.7 \\
33\% mixture & 1 & $ 29.7^{+7.1}_{-6.2} $ & 40 & 9 & 18.0 \\
66\% mixture & 3 & $ 52.7^{+5.9}_{-5.5} $ & 195 & 29 & 51.9 \\
modern & 5 & $ 103.9^{+6.7}_{-6.3} $ & 510 & 35 & 77.1 \\
\hline
\end{tabular}
\end{table}

\clearpage
\newpage
\mbox{~}

\begin{table}[h]
\centering
\caption{Overview of ArTTA measurements during ice project. The reported concentrations are the most probable values and the $1\sigma$ confidence interval is given. Several measurements are grouped together and use the mean of the included references. The highly enriched references and samples measured in throughput configuration do not suffer from significant background contribution. Before each sample analysis, the apparatus was flushed for $2$~hours with krypton. Independent measurements have confirmed that with that procedure the cross sample contamination is minimized.} 
\begin{tabular}{lrrrrrrr}
date & sample & function & concentration & total atoms detected & estimated background & time [h] & volume [mL STP]\\
\hline
04/17/18 & 10x enriched & reference & $ - $ & 94 & - & 1.83 & 1.9 \\
04/17/18 & 10x enriched & reference & $ - $ & 181 & - & 3.16 & 2.1 \\
04/17/18 & bottle argon & known sample & $108.7^{+14.1}_{-12.8}$ & 83 & - & 13.16 & throughput \\
04/18/18 & 10x enriched & reference & $ - $ & 109 & - & 2.00 & 2.0 \\
04/18/18 & Schaufelferner A & sample & $60.8^{+8.9}_{-8.0}$ & 92 & 13 & 22.30 & 1.7 \\
04/19/18 & 10x enriched & reference & $ - $ & 125 & - & 2.16 & 1.4 \\
\hline
04/24/18 & 10x enriched & reference & $ - $ & 94 & - & 2.16 & 1.6 \\
04/24/18 & dead & background & $ - $ & 11 & - & 15.65 & 1.2 \\
04/25/18 & 10x enriched & reference & $ - $ & 85 & - & 2.00 & 2.1 \\
04/25/18 & 10x enriched & reference & $ - $ & 118 & - & 2.50 & 1.5 \\
04/25/18 & dead & background & $ - $ & 16 & - & 19.99 & 1.5 \\
04/26/18 & 10x enriched & reference & $ - $ & 107 & - & 2.00 & 1.5 \\
04/26/18 & Schaufelferner B & sample & $60.0^{+10.1}_{-9.1} $ & 66 & 10 & 19.50 & 1.4 \\
04/27/18 & 10x enriched & reference & $ - $ & 87 & - & 2.00 & 1.5 \\
\hline
04/30/18 & 10x enriched & reference & $ - $ & 109 & - & 2.00 & 1.4 \\
05/01/18 & 10x enriched & reference & $ - $ & 121 & - & 2.00 & 1.4 \\
05/01/18 & bottle argon & known sample & $122.0^{+14.4}_{-13.2} $ & 102 & - & 14.66 & throughput\\
05/02/18 & 10x enriched & reference & $ - $ & 132 & - & 2.00 & 1.4 \\
05/02/18 & dead & background & $ - $ & 10 & - & 20.00 & 1.1 \\
05/03/18 & 10x enriched & reference & $ - $ & 101 & - & 2.00 & 1.5 \\
05/03/18 & Entensee & sample & $ 17.0^{+5.8}_{-5.2} $ & 33 & 14 & 20.00 & 1.3 \\
05/04/18 & 10x enriched & reference & $ - $ & 101 & - & 2.33 & 1.3 \\
\hline
05/06/18 & 10x enriched & reference & $ - $ & 98 & - & 2.16 & 0.5 \\
05/06/18 & Titlis 1-1 & sample & $25.7^{+9.2}_{-8.5} $ & 50 & 26 & 20.00 & 0.5 \\
05/07/18 & 10x enriched & reference & $ - $ & 96 & - & 2.00 & 0.5 \\
05/07/18 & dead & background & $ - $ & 21 & - & 20.00 & 0.5 \\
05/08/18 & 10x enriched & reference & $ - $ & 91 & - & 2.00 & 0.5 \\
05/08/18 & Titlis 1-2 & sample & $5.5^{+5.6}_{-5.3} $ & 31 & 25 & 20.00 & 0.6 \\
05/06/18 & 10x enriched & reference & $ - $ & 85 & - & 2.00 & 0.5 \\
\hline
\end{tabular}
\end{table}

\end{document}